# Thermodynamic and Thermoelectric Properties of CoFeYGe (Y= Ti, Cr) Quaternary Heusler Alloys: First Principle Calculations

*Raad Haleoot, Bothina Hamad*[*]

[1] *Microelectronics and Photonics graduate program, University of Arkansas, Fayetteville, AR 72701, USA*
[2] *Department of Physics at the College of Education, University of Mustansiriyah, Baghdad 10052, Iraq*
[3] *Department of Physics University of Arkansas, Fayetteville, AR 72701, USA*
[4] *Physics Department, The University of Jordan, Amman-11942, Jordan*

## Abstract

Utilizing a material in thermoelectric applications requires a mechanical, thermal, and lattice stability as well a high figure of merit (*ZT*). In this work, we present the structural, electronic, magnetic, mechanical, thermodynamic, dynamic, and thermoelectric properties of CoFeYGe (Y = Ti, Cr) quaternary Heusler compounds using the density functional theory (DFT). The calculated mechanical properties and phonon dispersions reveal that the structures of these compounds are stable. Both CoFeCrGe and CoFeTiGe compounds show a ferromagnetic and ferrimagnetic half-metallic behavior with band gaps of 0.41 and 0.38 eV, respectively. The lattice thermal conductivity ($\kappa_L$) exhibits low values that reach 3.01 W/(m.K) (3.47 W/(m.K)) for CoFeCrGe (CoFeTiGe) at 1100 K. The optical phonon modes have a large contribution of 60.2% (70.9 %) to $\kappa_L$ value for CoFeCrGe (CoFeTiGe). High *ZT* values of 0.71 and 0.65 were obtained for CoFeCrGe and CoFeTiGe, respectively. Based on our calculations, CoFeCrGe and CoFeTiGe combine both good spintronic and thermoelectric behaviors that may be used in spin injection applications.

# 1. Introduction

The demand is increasing to find new energy resources that are renewable, sustainable, and environmentally friendly. One of the solutions is by using thermoelectric materials that could convert waste heat into electricity and vice versa. The efficiency is measured using the figure of merit ($ZT = S^2\sigma T/(\kappa_e + \kappa_L)$), where $S$, $\sigma$, $T$, $\kappa_e$ and $\kappa_L$ are the Seebeck coefficient, electrical conductivity, absolute temperature, electronic and lattice thermal conductivities, respectively. Among the variety of thermoelectric materials, Heusler alloys showed novel properties such as full one electron spin polarization current,[1] low lattice thermal conductivity,[2] and high Seebeck coefficient.[3] Recent studies showed that Heusler compounds exhibit high ZT values as compared to the current state-of-the-art thermoelectric materials.[4,5] Hence, Heusler alloys become interesting materials for wide technology fields such as thermoelectric coupling and spintronic devices.[4,6,7] Full Heusler alloys consist of four interpenetrating face centered cubic sublattices with the chemical formula $X_2YZ$ where X and Y are transition or rare-earth atoms and Z atom is the main group element. When X atom is replaced by another transition metal atom $X_0$ then the resultant is quaternary Heusler alloy with a chemical formula $XX_0YZ$.

Cobalt-based quaternary Heusler alloys are very interesting materials for spintronic and thermoelectric applications. Vajiheh *et al.* synthesized CoFeMnZ (Z = Al, Ga, Si, Ge) compounds, which exhibit high Curie temperatures above 550 K and *ab initio* electronic structure calculations predicted these compounds as half-metallic ferromagnets where one spin channel is metallic, while the other spin channel exhibits a semiconducting behavior.[8] *Ab initio* electronic structure calculations showed that CoCrFeGe is dynamically stable because all phonon modes show positive frequencies.[9] Enamullah *et al.* investigated the effect of defects and change of pressure on the electronic and magnetic properties of CoFeCrGe and CoMnCrAl, where CoFeCrGe alloy retained its half-metallicity for a limited range of pressure.[10] A theoretical and experimental study by Enamullah *et al.* shows that CoFeCrGe alloy crystallizes in a Y-type structure and exhibits a half-metallic ferromagnetic behavior with high Curie temperature (866 K).[11] Density functional theory predicted that XX$_0$YZ (X = Co, Ni, X$_0$ = Fe; Y = Ti; Z = Si, Ge, As) quaternary Heusler alloys are half-metallic.[12] In addition, Zhang et al. investigated the effect of the lattice distortion on the half-metallicity of CoFeTiZ (Z = Ge, Si, Sn) where CoFeTiGe retained its half-metallic behavior for a large change in lattice constant.[13] Low lattice thermal conductivity and good Seebeck coefficient is predicted for CoFeCrZ (Z = Si, As, Sb) by Bhat *et al.*.[14] Density functional theory calculations predicted that CoFeCrAs has a Seebeck coefficient and a power factor ($S^2\sigma$) of -40 µV/K and 18×10$^{14}$ µWcm$^{-1}$K$^{-2}$s$^{-1}$, respectively, at room temperature.[15]

We systematically calculated the structural, elastic, electronic, magnetic, thermodynamic, lattice dynamic, and thermoelectric properties of CoFeYGe (Y = Ti, Cr) quaternary Heusler compounds by employing first principles calculations. These compounds are found to exhibit good spintronic and thermoelectric properties. Such properties are promising for applications in spin injection using the spin polarized thermoelectric current. The method of calculations is presented in section 2, results and discussions are presented in section 3, and the conclusion is given in section 4. The choice of a heavy metalloid atom is expected to reduce the phonon lattice thermal conductivity, while retaining the low value of electrical resistivity and large value of Seebeck coefficient.[16]

## 2. Method of calculations

The structural, electronic and magnetic properties of CoFeYGe (Y = Ti, Cr) are investigated using DFT calculations as implemented in VASP code[17] and the projector-augmented wave method.[18] The exchange-correlation potential was treated using the generalized-gradient approximation of Perdew-Burke-Ernzerhof (GGA-PBE).[19] Full relaxation of cell volume, shape and atomic positions are performed. To ensure high accuracy results, the system energy and force convergence criteria are set to be 1×10$^{-7}$ eV and 1×10$^{-6}$ eV Å$^{-1}$, respectively. The plane-wave cut-off energy was set to 500 eV and Γ-centered k-point mesh of 24×24×24 was used in the Brillouin zone integrations. Phonon properties and second-order harmonic interatomic force constants are calculated using Phonopy package with 4×4×4 supercell.[20] The thirdorder.py module of the ShengBTE package[21] was used to calculate third-order harmonic interatomic force constants where a 4×4×4 supercell with seventh nearest neighbor interactions were taken into account. The lattice thermal conductivity was calculated using Boltzmann transport theory for phonons as implemented in the ShengBTE with a dense q mesh 28×28×28, which is the converged value. Further increase of the q mesh lead to insignificant effect of the value of the thermal conductivity (less than 3% by doubling the q mesh). The thermoelectric properties of the alloys are computed using Boltzmann transport theory as implemented in the BoltzTrap program.[22]

## 3. Results and discussions

The structural, electronic, magnetic, thermodynamic, dynamic, and thermoelectric properties of CoFeYGe (Y = Ti, Cr) are presented in the following subsections.

### 3.1 Structural properties

Quaternary Heusler compounds crystallize in the LiMgPdSn structure with three possible types of nondegenerate atomic configurations named as Y-type structures,[23] see **Table 1**. The Y-type-I structure with a space group $F\bar{4}3m$ (#216) is found to be the most stable geometrical configuration of CoFeYGe (Y = Ti, Cr) compounds. The calculated lattice parameters are found to be in good agreement with previous theoretical[10,13] and experimental[11] results as shown in **Table 2**.

**Table 1.** The three structural configurations types of CoFeYGe (Y= Ti, Cr) quaternary Heusler alloys. The Wyckoff positions of 4a, 4b, 4c, and 4d are (0, 0, 0, 0), (½, ½, ½, ½), (¼, ¼, ¼, ¼), and (¾, ¾, ¾, ¾), respectively.

| Type | 4a | 4b | 4c | 4d |
|---|---|---|---|---|
| Y-type-I | Co | Fe | Y | Ge |
| Y-type-II | Co | Y | Fe | Ge |
| Y-type-III | Fe | Y | Co | Ge |

Investigating the mechanical properties of a material is crucial before adopting it in technological applications. Therefore, the mechanical stability is checked for these compounds using the Born-Haung criteria.[24] There are three independent elastic constants for the cubic structure, namely, $C_{11}$, $C_{12}$, and $C_{44}$. The conditions of mechanical stability are as follows:

$$(C_{11} - C_{12})/2 > 0, \ (C_{11} + 2C_{12})/3 > 0, \ C_{44} > 0. \tag{1}$$

The elastic constants are calculated using the stress-strain relationship.[25] It is clear from Table 2 that the elastic constants satisfy the stability conditions, which confirm that both compounds CoFeYGe (Y = Ti, Cr) are mechanically stable in Y-type-I structure. The elastic constants used to evaluate the elastic moduli such as bulk modulus $B$, and shear modulus $G$ are calculated using the following relations[26,27]

$$B = (C_{11} + 2C_{12})/3 \tag{2}$$

$$G_R = (5C_{44}(C_{11} - C_{12}))/4C_{44} + (C_{11} - C_{12}) \tag{3}$$

$$G_V = (C_{11} - C_{12} + 3C_{44})/5 \tag{4}$$

Where $G_R$ and $G_V$ are Reuss's and Voigt's shear moduli, respectively. Voigt-Reuss-Hill approximation is used to estimate the arithmetic average of the shear modulus using the following equation[28]

$$G = (G_V + G_R)/2 \tag{5}$$

In addition, Young's modulus $E$, Poisson's ratio $v$, and anisotropy factor $A$ are calculated as follows:

$$E = 9BG/(3B + G) \tag{6}$$

$$v = (3B - 2G)/2(3B + G) \tag{7}$$

$$A = 2C_{44}/(C_{11} - C_{12}) \tag{8}$$

The calculated bulk moduli, Young's moduli, isotropic shear moduli, Poisson's ratios, anisotropy factor, and Pugh's *B/G* ratio are given in Table 2. The bulk modulus *B* measures the resistance of a material to compression, whereas Young's modulus *E* provides information about its stiffness where the higher value corresponds to a stiffer material. The calculated bulk modulus of CoFeCrGe compound (212.64 GPa) is found to be in agreement with the calculated value (215.17 GPa) by Enamullah *et al.*.[10] The results show that *B* and *E* values of CoFeCrGe are higher than those of CoFeTiGe. However, the shear modulus of CoFeTiGe is found to be higher than that of CoFeCrGe.

This indicates that the lateral deformation due to the applied shear force is higher in the case of CoFeTiGe than that of CoFeCrGe. The ratio of the transverse contraction to the longitudinal extension in the direction of elastic loading is defined as Poisson's ratio ($v$), where the lower limit of $v$ for most metals is about 0.25.[29] The calculated $v$ values of CoFeTiGe and CoFeCrGe are 0.337 and 0.336, respectively, which are comparable to those of CoRuFeGe (0.336) and CoRuFeSn (0.365) quaternary Heusler alloys.[30] The values of the anisotropy factor $A$ of CoFeYGe (Y= Ti, Cr) compounds is found to deviate from unity, which indicates an anisotropic elastic behavior. Based on the Pugh's criteria, the value of $B/G$ provides information about ductile (brittle) nature of materials.[31] For $B/G$ higher (lower) than 1.75, the material behaves in a ductile (brittle) manner. The $B/G$ ratios of CoFeYGe (Y = Ti, Cr) are found to be higher than 1.75, see Table 2, which reveals the ductile behavior of these compounds. The melting temperature of the compounds is calculated using the following empirical relation[32]

$$T_{melt} = 607 + 9.3 \times B \pm 555 \tag{9}$$

From the above equation, the melting temperature of CoFeCrGe is higher than CoFeTiGe due to the higher bulk modulus of the former than the latter compound.

**Table 2.** The calculated lattice constant (a, in [Å]), elastic constant ($C_{ij}$, in [GPa]), bulk modulus ($B$, in [GPa]), Young's modulus ($E$, in [GPa]), isotropic shear modulus ($G$, in [GPa]), Poisson's ratios ($v$), anisotropy factor ($A$), Pugh's ratio ($B/G$), and melting temperature ($T_{melt}$, [K]) of CoFeYGe (Y=Ti, Cr) compounds.

| Type | a | $C_{11}$ | $C_{12}$ | $C_{44}$ | B | E | G | v | A | B/G | $T_{melt}$ |
|---|---|---|---|---|---|---|---|---|---|---|---|
| CoFeCrGe | 5.71 | 354.08 | 141.92 | 66.99 | 212.64 | 171.26 | 63.23 | 0.366 | 0.62 | 3.36 | 2584 |
|  | 5.77[a)] |  |  |  |  | 215.17[c)] |  |  |  |  |  |
|  | 5.71[b)] |  |  |  |  |  |  |  |  |  |  |
| CoFeTiGe | 5.81 | 328.45 | 138.62 | 71.10 | 201.89 | 197.81 | 74.05 | 0.337 | 0.76 | 2.72 | 2484 |
|  | 5.81[d)] |  |  |  |  |  |  |  |  |  |  |

[a)]Ref.[11] Exp.; [b)]Ref.[10] Theo.; [c)]Ref.[10] Theo.; [d)]Ref.[13] Theo.

**3.2 Electronic and magnetic properties**

The spin-resolved band structure of CoFeYGe (Y = Ti, Cr) alloys are presented in **Figure 1**. The results show that CoFeYGe (Y = Ti, Cr) alloys are half-metallic, where the majority and minority spin channels possess metallic and semiconducting behaviors, respectively. Therefore, the two compounds show a 100% spin polarization, which makes them promising candidates for spintronic applications.[33] Both CoFeCrGe and CoFeTiGe compounds exhibit indirect band gaps along Γ-X high symmetry line of 0.412 and 0.383 eV, in agreement with previous calculations of 0.481 and 0.424 eV, respectively.[11,13] The minority valence band is close to the Fermi energy at Γ high symmetry point for both compounds. The presence of flat energy levels in the conduction bands along Γ-X and highly dispersive bands along other directions could be a signature of high thermoelectric power factor.[34]

**Table 3** presents the calculated total and local magnetic moments for CoFeYGe (Y= Ti, Cr) as well as reported experimental and theoretical results for comparison. The results show that both CoFeCrGe and CoFeTiGe are magnetic materials with total magnetic moments of 3.0 $\mu_B$ and 1.0 $\mu_B$, respectively. The integer values of the total magnetic moments obey Slater-Pauling rule for half metallic magnetic materials as follows[35]

$$M_{tot} = (Z_{tot} - 24)\mu_B \qquad (10)$$

Where $M_{tot}$ and $Z_{tot}$ are the total magnetic moment and the number of total valence electrons, respectively.

The local magnetic moments of Co, Fe, and Cr in the case of CoFeCrGe alloy are ferromagnetically coupled, where the maximum contribution comes from Cr atom that carries the highest magnetic moment (1.835 $\mu_B$). However, the local magnetic moment of Ti atom (-0.226 $\mu_B$) is coupled antiferromagnetically with Co and Fe atoms in the case of CoFeTiGe alloy. For both compounds, the Ge atom showed a negligible value of the induced magnetic moment.

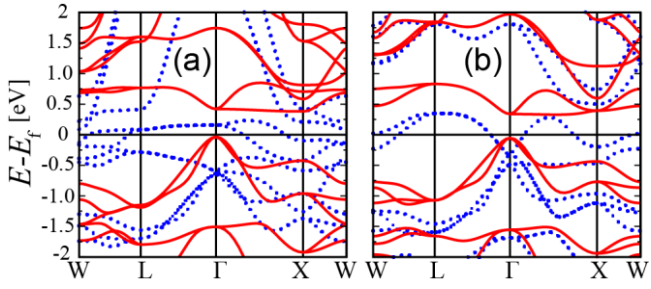

**Figure 1.** The spin-polarized band structure of a) CoFeCrGe and b) CoFeTiGe in the first Brillouin zone. Spin-up and spin-down are denoted by dotted and solid lines, respectively.

**Table 3.** The calculated half-metallic energy gap ($E_g$), total and local magnetic moments CoFeYGe (Y=Ti, Cr) compounds.

| Compound | $E_g$ [eV] | $m_{Co}$ [$\mu_B$] | $m_{Fe}$ [$\mu_B$] | $m_Y$ [$\mu_B$] | $m_{Ge}$ [$\mu_B$] | $m_{total}$ [$\mu_B$] |
|---|---|---|---|---|---|---|
| CoFeCrGe | 0.412 | 1.01 | 0.19 | 1.83 | -0.04 | 3.00 |
|  | 0.481[a] |  |  |  |  | 3.00[a] |
| CoFeTiGe | 0.383 | 0.55 | 0.69 | -0.23 | -0.01 | 1.00 |
|  | 0.424[b] | 0.56[b] | 0.72[b] | -0.32[b] | 0.04[b] | 1.00[b] |

[a]Ref.[11] Theo.; [b]Ref.[13] Theo. and Exp.

### 3.3 Thermodynamic properties

Utilizing CoFeYGe (Y= Ti, Cr) quaternary Heusler compounds as thermoelectric materials requires an understanding of their thermodynamic properties. The quasi-harmonic approximation (QHA) works well for the temperatures below the melting point.[20] Therefore, the QHA is used as implemented in Phonopy-QHA package[20] to investigate the effect of temperature on various thermodynamic parameters, such as the thermal expansion coefficient and the variation of volume and bulk modulus with temperature.[20] These parameters are calculated in the temperature range from 0 to 1200 K at a constant pressure of 0 GPa. **Figure 2**a shows the normalized volume $V/V_0$ as a function of temperature up to 1200 K, where V is the volume of the 4×4×4 supercell at temperature T and $V_0$ is its volume at 0 K. This figure shows that the volume of the unit cell is increasing linearly as a function of temperature beyond 50 K for both compounds. The bulk modulus is also investigated to examine the effect of temperature on the hardness of the compounds, see Figure 2b. At low temperature, the variation of the bulk modulus is negligible due to the insignificant change of the volume in this region. As the temperature increases, the bulk modulus of CoFeCrGe compound exhibits a higher negative slope than CoFeTiGe alloy.

Another important thermal parameter is the thermal expansion coefficient $\alpha$ as depicted in Figure 2c. At low temperatures, the thermal expansion coefficient starts with negligible values then increases rapidly up to 200K. At higher temperatures, $\alpha$ increases linearly as a function of temperature for the case of CoFeCrGe alloy, while it approaches a constant value for CoFeTiGe. At 300 K, the value of the thermal expansion coefficient equals $8.14 \times 10^{-6}$ $K^{-1}$ and $7.86 \times 10^{-6}$ $K^{-1}$ for CoFeCrGe and CoFeTiGe, respectively, which are lower than that of FeCrRuSi quaternary Heusler alloy ($5.97 \times 10^{-5}$ $K^{-1}$).[36]

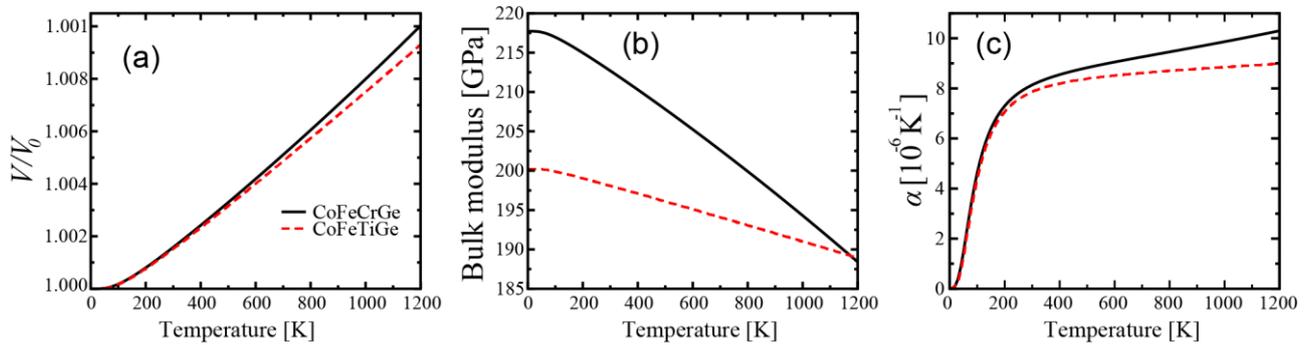

**Figure 2.** a) The normalized volume $V/V_0$, b) the variation of bulk modulus, and c) the thermal expansion coefficient as a function of temperature of CoFeYGe (Y = Ti, Cr) compounds.

**3.4 Lattice dynamics**

The heat capacity $C_v$ is an essential thermal property of a material, which is calculated using the phonon density of states. Figure S1 in the Supporting Information depicts the variation of heat capacity with temperature. At low temperatures, the variation of $C_v$ is proportional to $T^3$. However, at higher temperatures, $C_v$ converges to the classical limit of Dulong and Petit that assumes a constant specific heat for solids. At 300K, CoFeCrGe alloy exhibits a slightly higher $C_v$ value (92.19 J mole$^{-1}$ K$^{-1}$) than CoFeTiGe (91.93 J mole$^{-1}$ K$^{-1}$) due to the higher molar mass of Cr as compared to Ti atom.

The phonon dispersion curves of CoFeYGe (Y = Ti, Cr) along high symmetry points in the first Brillouin zone, including the non-analytical term correction, are shown in **Figure 3**a,d. The phonon dispersion curves show twelve branches, which are the sum of the three acoustic and nine optical modes due to the presence of four atoms in the primitive cell. The three acoustic modes consist of one longitudinal (LA) and two transversal (TA) branches. From Figure 3a,d, one can see only two acoustic branches along Γ-X and Γ-L due to the degeneracy of the transverse branches by symmetry. However, L-W path shows two TA branches in addition to the LA branch. Similar dispersion relations were reported for YPdSb half Heusler alloy.[37] The absence of negative phonon frequencies in the phonon dispersion curves indicate the mechanical stability of CoFeYGe (Y = Ti, Cr) quaternary Heusler compounds. The partial density of states (PDOS) of phonons is shown in Figure 3b,e. The vibration of the heavy Ge and Co atoms in CoFeYGe (Y = Ti, Cr) compounds are found to dominate the low frequency region, while Y atoms show higher contributions in the high-frequency region. The group velocity ($v_g = \partial \omega(k)/\partial k$) is defined as the slope of the dispersion relation, which is important to determine the lattice thermal conductivity. Figures 3c,f depict the group velocity of CoFeYGe (Y = Ti, Cr) quaternary Heusler compounds. From these figures, it is clear that the optical modes show a significant contribution to the group velocity, which comes from the non-flat optical phonon branches as shown in Figure 3a,d.

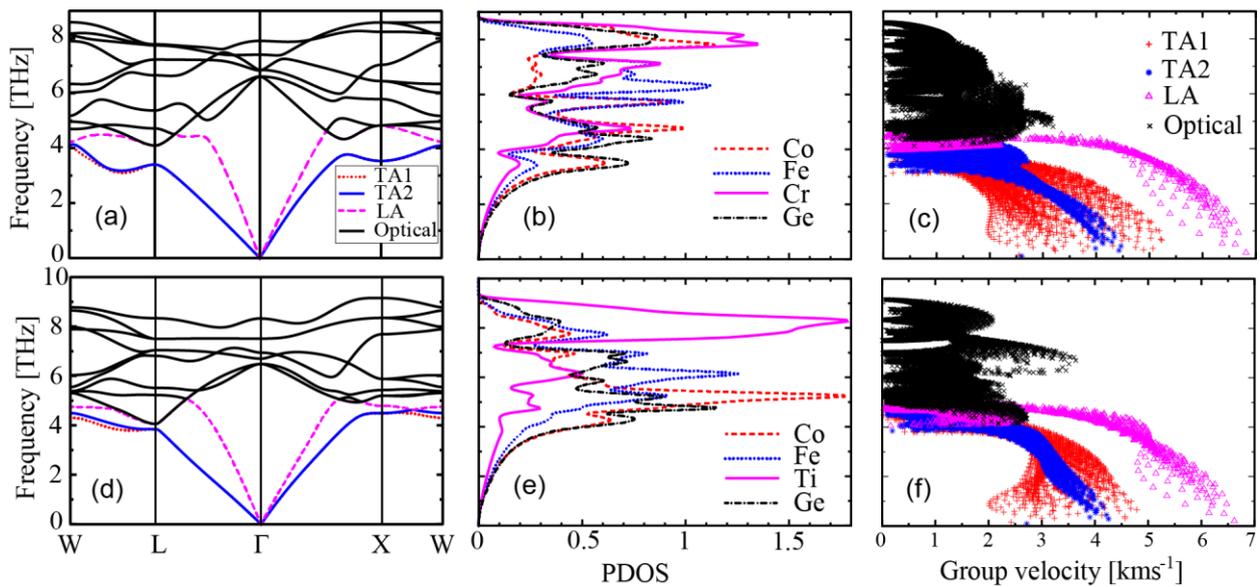

**Figure 3.** The phonon dispersion relation, partial density of states and group velocity of a,b,c) CoFeCrGe and d,e,f) CoFeTiGe.

The lattice thermal conductivity ($\kappa_L$) is an essential quantity that plays a vital role in designing devices such as thermoelectric generators. It can be calculated using the following equation[21]

$$k_l^{\alpha\beta} = \frac{1}{k_B T^2 \Omega N} \sum_\lambda f_0(f_0 + 1)(\hbar\omega_\lambda)^2 v_\lambda^\alpha F_\lambda^\beta \qquad (11)$$

Where $\Omega$, $N$, $f_0$, $\omega_\lambda$, $v_\lambda$ are the volume of the unit cell, a number of q-points, phonon distribution function at thermal equilibrium, the angular frequency of phonon mode $\lambda$, group velocity along the $\alpha$ direction, and scattering along direction $\beta$. The lattice thermal conductivity is found to decrease as a function of temperature for CoFeYGe (Y = Ti, Cr) quaternary Heusler compounds, see **Figure 4**a. The values of $\kappa_L$ at room temperature (300K) are found to be 11.01 and 12.26 Wm$^{-1}$K$^{-1}$ for CoFeCrGe and CoTiFeGe, respectively. These values are close to the well-known thermoelectric material CoSb$_3$ (10 Wm$^{-1}$K$^{-1}$).[38,39] By elevating the temperature to 1100 K, $\kappa_L$ decreases to 3.01 and 3.47 Wm$^{-1}$K$^{-1}$ for CoFeCrGe and CoFeTiGe, respectively. This can be ascribed to the increase in the scattering rate of excited phonons that leads to a reduction in the mean free path. The $\kappa_L$ values of CoFeTiGe are found to be slightly higher than those of CoFeCrGe due to the lighter atomic mass of Ti than that of Cr. The absence of the gap between the highest acoustic modes and the lowest optical modes in Figure 3a,d could be the reason for the low $\kappa_L$.[40,41] Such gaps, if existing, are related to the difference in the atomic mass of the constituent elements in the compound.[42] This difference in the atomic masses means a lower probability of the three-phonon scattering, which leads to longer relaxation times that corresponds to a larger thermal conductivity.[43]

Investigating the contribution of the different phonon modes to the overall lattice thermal conductivity provides an insight for the design of the thermoelectric materials. In most cases, the optical modes have a negligible contribution to the thermal conductivity as compared to the acoustic modes due to its low group velocity. In contrast, CoFeYGe (Y = Ti, Cr) compounds show unconventional results where the contribution of the optical modes is 60.2% and 70.9 % at 1100K for CoFeCrGe and CoFeTiGe, respectively as shown in Figure S2 in the Supporting Information. This is not surprising, because the group velocity of the optical branches is significant due to the non-flat optical phonon branches as shown in Figure 3a,d. This is consistent with previous studies of several materials.[44–46] As the optical modes are dominated by the contribution of Cr (Ti) atoms, vacancies at these sites could lead to a significant reduction in the thermal conductivity and an increase in *ZT* values.

To investigate the effect of the grain size on the thermal conductivity, we plot the normalized accumulated thermal conductivity as a function of the mean free path (MFP) for different temperatures, see Figure 4b,c. From these figures, one can see that the size effect significantly influences the thermal conductivity. As the temperature increases, the mean free path of the phonons decreases, which leads to a stronger anharmonic phonon scattering. The reduction of the lattice thermal conductivity by 50% at 300 K requires MFP values of 12.75 nm and 12.05 nm for CoFeCrGe and CoFeTiGe, respectively. These MFP values are even lower at higher temperatures such as 1100K (3.00 nm and 3.18 nm for CoFeCrGe and CoFeTiGe, respectively). However, it is relatively challenging to reach this size by nanoengineering.

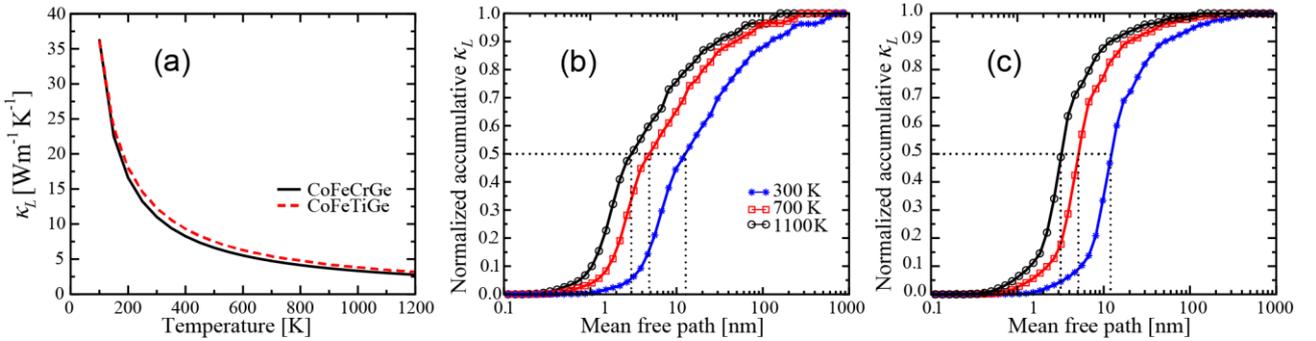

**Figure 4.** a) The temperature dependence of the lattice thermal conductivity of CoFeYGe (Y=Ti, Cr) compounds. The normalized accumulated lattice thermal conductivity as a function of the mean free path at different temperatures for b) CoFeCrGe and (c) CoFeTiGe.

**3.5 Thermoelectric properties**

Boltzmann's transport theory is employed to investigate the electronic transport properties of CoFeYGe (Y = Ti, Cr) using the constant relaxation time approximation as implemented in BoltzTraP code.[22] Based on this approximation, the transport coefficient tensors, electrical conductivity ($\sigma$), electronic thermal conductivity ($k_e$), and Seebeck coefficient (*S*) are given as[47,48]

$$\sigma_{\alpha\beta}(T;\mu) = \frac{1}{\Omega} \int \sigma_{\alpha\beta}(\varepsilon) \left[ -\frac{\partial f_\mu(T;\epsilon)}{\partial \varepsilon} \right] d\varepsilon \qquad (12)$$

$$k_{\alpha\beta}(T;\mu) = \frac{1}{e^2 T\Omega} \int \sigma_{\alpha\beta}(\varepsilon)(\varepsilon-\mu)^2 \left[-\frac{\partial f_\mu(T;\varepsilon)}{\partial \varepsilon}\right] d\varepsilon \qquad (13)$$

$$S_{\alpha\beta}(T;\mu) = \frac{1}{eT\Omega\sigma_{\alpha\beta}(T;\mu)} \int \sigma_{\alpha\beta}(\varepsilon)(\varepsilon-\mu) \left[-\frac{\partial f_\mu(T;\varepsilon)}{\partial \varepsilon}\right] d\varepsilon \qquad (14)$$

where $\alpha$ and $\beta$ are tensor indices, $\mu$, $\Omega$, $\varepsilon$, $e$, and $f$ are the chemical potential, volume of unit cell, band energy, Fermi level of carriers, electron charge, and the carrier Fermi-Dirac distribution function, respectively. In this work, the relaxation time $\tau$ is assumed to be $10^{-14}$ s, which is a typical value for semiconductors.[49,50] The rigid band approximation is used to evaluate the thermoelectric properties of CoFeYGe (Y = Ti, Cr) compounds. This approximation assumes that the band structure of the thermoelectric material does not change by doping, whereas the Fermi level is shifted towards the conduction band (valence band) in *n*-type (*p*-type) doping. Both CoFeCrGe and CoFeTiGe alloys exhibit a half-metallic behavior with a narrow minority band gaps of 0.412 eV and 0.383 eV, respectively, which indicate good thermoelectric properties.[51] The transport properties of CoFeYGe (Y = Ti, Cr) quaternary Heusler compounds are investigated as a function of the chemical potential for specific different temperatures (300, 700, and 1100 K).

For half-metallic materials, total Seebeck coefficient (a measure of the ability of a material to produce voltage due to a temperature difference) can be evaluated by using the two-current model as follows[52]

$$S = (S_\uparrow \sigma_\uparrow + S_\downarrow \sigma_\downarrow)/\sigma_{total} \qquad (15)$$

where $S_\uparrow$ ($S_\downarrow$) and $\sigma_\uparrow$ ($\sigma_\downarrow$) are the Seebeck coefficients and electrical conductivities for the spin-up (spin-down) channel, respectively. The $\sigma_{total} = (\sigma_\uparrow + \sigma_\downarrow)$ represents the total electrical conductivity. Figure S3 in the Supporting Information shows the total and the spin dependent Seebeck coefficient as a function of the chemical potential at 300 K. The spin-down channel for CoFeYGe (Y = Ti, Cr) exhibits high Seebeck coefficient, however, the total value is lower than that of the spin-down channel. To understand the behavior of the total Seebeck coefficient, the total and spin electrical conductivity are depicted in Figure S4 in the Supporting Information. While $\sigma_\uparrow$ exhibits high values, $\sigma_\downarrow$ shows negligible contribution in the region around the $\mu\text{-}E_f = 0$ eV.

**Figure 5**a,b are plotted to investigate the change of the total Seebeck coefficient as a function of the chemical potential with different temperatures. The Seebeck coefficient varies with temperature, where CoFeTiGe alloy shows lower *n*- and *p*-type values than CoFeCrGe that has peak values of -81.2 µV/K and 43.1 µV/K at 1100 K for the *n*- and *p*-type, respectively. The total electrical conductivity as a function of chemical potential is shown in Figure 5c,d at different temperatures. From the figure, the *n*-type shows larger increase in values of the electrical conductivity than the *p*-type. This increase is larger in the case of CoFeCrGe than CoFeTiGe. It is also noticed that $\sigma$-$\mu$ curves are less sensitive to the temperature change. The electronic thermal conductivity is related to the electronic conductivity by Wiedemann–Franz law ($\kappa_e = L\sigma T$), where $L$ is the Lorentz number. Therefore, the trend of total $\kappa_e$ as a function of $\mu$ in Figure 5e,f is similar to $\sigma$ in Figure 5c,d. However, unlike $\sigma$, the $\kappa_e$ is temperature dependent, which increases at elevated temperatures.

The performance of thermoelectric material is represented by the figure of merit (*ZT*), which is shown in Figure 5g,h as a function of chemical potential at different temperatures. The *ZT* values of CoFeYGe (Y = Ti, Cr) quaternary Heusler compounds are found to increase at higher temperatures. There are two peaks of *ZT* for each compound one for *n*-type and the other for *p*-type. The *n*-type of CoFeCrGe compound exhibits a maximum *ZT* of 0.71 at 1100 K, while the *p*-type has a peak at 0.46. In the case of CoFeTiGe compound, the two peaks of *ZT* at 1100 K have close values of 0.65 and 0.61 as *p*-type and *n*-type, respectively. Therefore, this material can work as a *p*-type or *n*-type with the same thermoelectric efficiency. It is worth mentioning that *ZT* value of CoFeTiGe is negligible at $\mu$-$E_f$ = 0 eV, unlike CoFeCrGe that has *ZT* of 0.28 at $\mu$-$E_f$ = 0 eV. This means that the latter compound has good thermoelectric properties even without doping. For both compounds, the chemical potential that corresponds to the peak value of *ZT* is shifted by varying the temperature. Therefore, it is important to choose the optimal carrier concentration for a specific temperature to achieve the highest *ZT* value.

The thermoelectric efficiency of a material can be enhanced by nanoengineering technology.[53–55] It was possible to synthesized CoFeCrGe with a grain size of 5 nm by Jin *et al.*.[56]

At MFP of 5 nm, the lattice thermal conductivity decreases to 1.88 Wm$^{-1}$K$^{-1}$ at room temperature (Figure S5 in the Supporting Information), which corresponds to an increase in the ZT from its bulk value of 0.14 to 0.4, which means a higher TE efficiency.

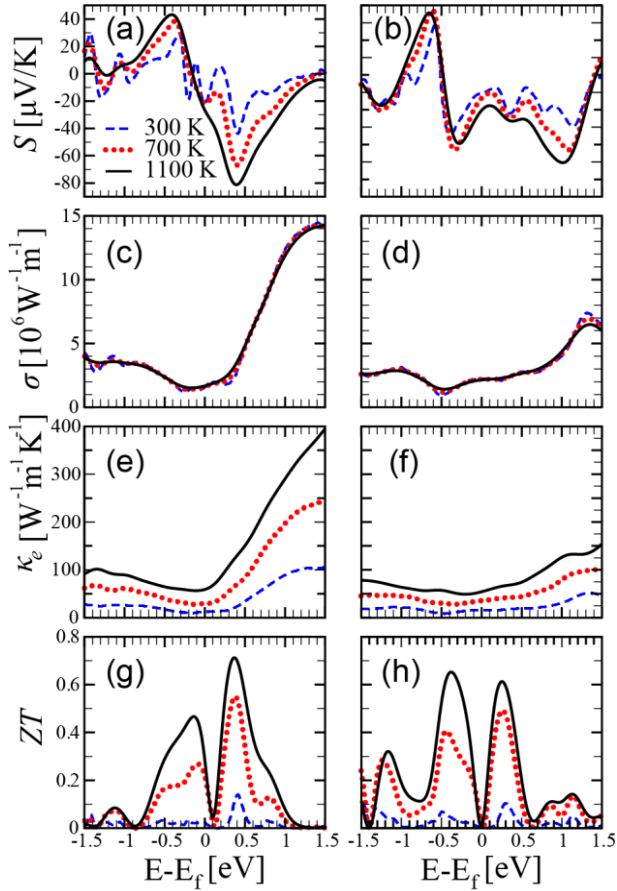

**Figure 5.** The Seebeck coefficient ($S$), electrical conductivity ($\sigma$), electronic thermal conductivity ($\kappa_e$), and the figure of merit ($ZT$) as a function of the chemical potential at different temperatures for CoFeCrGe (left column) and CoFeTiGe (right column).

## 4. Conclusion

The electronic, magnetic, and thermoelectric properties of CoFeYGe (Y = Ti, Cr) quaternary Heusler compounds are investigated using DFT calculations and the semiclassical transport theory. The results indicate a half-metallic behavior of CoFeCrGe and CoFeTiGe quaternary Heusler compounds with band gaps of 0.412 and 0.383 eV, respectively, in the minority spin channels. The compounds were found to be magnetic with total magnetic moments are 3.00$\mu_B$ and 1.00$\mu_B$ for CoFeCrGe and CoFeTiGe, respectively. The calculated values of the lattice thermal conductivity are 11.01(3.01) and 12.26 (3.47) Wm$^{-1}$K$^{-1}$ for CoFeCrGe and CoTiFeGe, respectively at 300 K (1100 K). The lattice thermal conductivity is found to be dominated by the optical rather than the acoustic phonon modes. The compounds show promising thermoelectric properties at 1100 K with appreciable *ZT* values of 0.71 and 0.65 for the *n*-type and the *p*-type of CoFeCrGe and CoFeTiGe, respectively. The interesting magnetic half metallicity as well as the good thermoelectric properties suggest these compounds as good candidates for applications using spin polarized thermoelectric current.

**Acknowledgments**
R. H. was funded by The Higher Committee for Education Development of Iraq. This work was supported by Arkansas NASA EPSCOR Research Infrastructure Development (RID) grant number 002276-00001A.

**Conflicts of interest**
The authors confirm that there is no conflict of interest with any other party regarding the material discussed in the manuscript.

References

[1]  A. Kundu, S. Ghosh, R. Banerjee, S. Ghosh, B. Sanyal, *Sci. Rep.* **2017**, *7*, 1803.
[2]  J. He, M. Amsler, Y. Xia, S. S. Naghavi, V. I. Hegde, S. Hao, S. Goedecker, V. Ozoliņš, C. Wolverton, *Phys. Rev. Lett.* **2016**, *117*, 046602.
[3]  R. Haleoot, B. Hamad, *J. Electron. Mater.* **2019**, *48*, 1164.
[4]  H. Zhu, R. He, J. Mao, Q. Zhu, C. Li, J. Sun, W. Ren, Y. Wang, Z. Liu, Z. Tang, A. Sotnikov, Z. Wang, D. Broido, D. J. Singh, G. Chen, K. Nielsch, Z. Ren, *Nat. Commun.* **2018**, *9*, 2497.
[5]  H. Zhu, J. Mao, Y. Li, J. Sun, Y. Wang, Q. Zhu, G. Li, Q. Song, J. Zhou, Y. Fu, R. He, T. Tong, Z. Liu, W. Ren, L. You, Z. Wang, J. Luo, A. Sotnikov, J. Bao, K. Nielsch, G. Chen, D. J. Singh, Z. Ren, *Nat. Commun.* **2019**, *10*, 270.
[6]  V. Alijani, J. Winterlik, G. H. Fecher, S. S. Naghavi, C. Felser, *Phys. Rev. B* **2011**, *83*, 184428.
[7]  S. Chen, Z. Ren, *Mater. Today* **2013**, *16*, 387.
[8]  V. Alijani, S. Ouardi, G. H. Fecher, J. Winterlik, S. S. Naghavi, X. Kozina, G. Stryganyuk, C. Felser, E. Ikenaga, Y. Yamashita, S. Ueda, K. Kobayashi, *Phys. Rev. B* **2011**, *84*, 224416.
[9]  A. İyigör, ş. Uğur, *Philos. Mag. Lett.* **2014**, *94*, 708.
[10] Enamullah, D. D. Johnson, K. G. Suresh, A. Alam, *Phys. Rev. B* **2016**, *94*, 184102.
[11] Enamullah, Y. Venkateswara, S. Gupta, M. R. Varma, P. Singh, K. G. Suresh, A. Alam, *Phys. Rev. B* **2015**, *92*, 224413.
[12] A. Amudhavalli, R. Rajeswarapalanichamy, K. Iyakutti, *J. Magn. Magn. Mater.* **2017**, *441*, 21.
[13] Y. J. Zhang, Z. H. Liu, G. T. Li, X. Q. Ma, G. D. Liu, *J. Alloys Compd.* **2014**, *616*, 449.
[14] T. M. Bhat, D. C. Gupta, *J. Phys. Chem. Solids* **2018**, *112*, 190.
[15] T. M. Bhat, D. C. Gupta, *J. Magn. Magn. Mater.* **2018**, *449*, 493.
[16] Y. Nishino, S. Deguchi, U. Mizutani, *Phys. Rev. B* **2006**, *74*, 115115.
[17] G. Kresse, D. Joubert, *Phys. Rev. B* **1999**, *59*, 1758.
[18] P. E. Blöchl, *Phys. Rev. B* **1994**, *50*, 17953.
[19] J. P. Perdew, K. Burke, M. Ernzerhof, *Phys. Rev. Lett.* **1996**, *77*, 3865.

[20] A. Togo, I. Tanaka, *Scr. Mater.* **2015**, *108*, 1.
[21] W. Li, J. Carrete, N. A. Katcho, N. Mingo, *Comput. Phys. Commun.* **2014**, *185*, 1747.
[22] G. K. H. Madsen, D. J. Singh, *Comput. Phys. Commun.* **2006**, *175*, 67.
[23] X. Dai, G. Liu, G. H. Fecher, C. Felser, Y. Li, H. Liu, *J. Appl. Phys.* **2009**, *105*, 07E901.
[24] Born, M, Huang, K, *Dynamical theory of crystal lattices*; Clarendon Press: Oxford, 1954.
[25] Y. Le Page, P. Saxe, *Phys. Rev. B* **2002**, *65*, 104104.
[26] A. Reuss, *ZAMM - Z. Für Angew. Math. Mech.* **1929**, *9*, 49.
[27] W. Voigt, *Lehrbuch der Kristallphysik*; Taubner, Leipzig, 1928.
[28] R. Hill, *Proc Phys Soc A* **1952**, *65*, 349.
[29] G. N. Greaves, A. L. Greer, R. S. Lakes, T. Rouxel, *Nat. Mater.* **2011**, *10*, 823.
[30] K. Benkaddour, A. Chahed, A. Amar, H. Rozale, A. Lakdja, O. Benhelal, A. Sayede, *J. Alloys Compd.* **2016**, *687*, 211.
[31] S. F. Pugh, *Lond. Edinb. Dublin Philos. Mag. J. Sci.* **1954**, *45*, 823.
[32] *Solid state physics simulations*; Johnston, I. D.; Consortium for Upper Level Physics Software, Eds.; Wiley: New York, 1996.
[33] I. I. Mazin, *Phys. Rev. Lett.* **1999**, *83*, 1427.
[34] D. I. Bilc, G. Hautier, D. Waroquiers, G.-M. Rignanese, P. Ghosez, *Phys. Rev. Lett.* **2015**, *114*, 136601.
[35] K. Özdoğan, E. Şaşıoğlu, I. Galanakis, *J. Appl. Phys.* **2013**, *113*, 193903.
[36] X. Wang, H. Khachai, R. Khenata, H. Yuan, L. Wang, W. Wang, A. Bouhemadou, L. Hao, X. Dai, R. Guo, G. Liu, Z. Cheng, *Sci. Rep.* **2017**, *7*, 16183.
[37] F. Kong, Y. Hu, H. Hou, Y. Liu, B. Wang, L. Wang, *J. Solid State Chem.* **2012**, *196*, 511.
[38] T. Caillat, A. Borshchevsky, J. -P. Fleurial, *J. Appl. Phys.* **1996**, *80*, 4442.
[39] D. T. Morelli, T. Caillat, J.-P. Fleurial, A. Borshchevsky, J. Vandersande, B. Chen, C. Uher, *Phys. Rev. B* **1995**, *51*, 9622.
[40] X. Wu, J. Lee, V. Varshney, J. L. Wohlwend, A. K. Roy, T. Luo, *Sci. Rep.* **2016**, *6*, 22504.
[41] X. Gu, R. Yang, *Appl. Phys. Lett.* **2014**, *105*, 131903.
[42] R. Guo, X. Wang, B. Huang, *Sci. Rep.* **2015**, *5*, 7806.
[43] T. Feng, X. Ruan, *J. Nanomater.* **2014**, *2014*, 1.
[44] M. D. Santia, N. Tandon, J. D. Albrecht, *Appl. Phys. Lett.* **2015**, *107*, 041907.
[45] S. Mukhopadhyay, L. Lindsay, D. J. Singh, *Sci. Rep.* **2016**, *6*, 37076.
[46] Z. Yan, S. Kumar, *Phys. Chem. Chem. Phys.* **2018**, *20*, 29236.
[47] J. Yang, H. Li, T. Wu, W. Zhang, L. Chen, J. Yang, *Adv. Funct. Mater.* **2008**, *18*, 2880.
[48] H. Kara, M. Upadhyay Kahaly, K. Özdoğan, *J. Alloys Compd.* **2018**, *735*, 950.
[49] R. Guo, X. Wang, Y. Kuang, B. Huang, *Phys. Rev. B* **2015**, *92*, 115202.
[50] P. Y. Yu, M. Cardona, *Fundamentals of Semiconductors*; Graduate Texts in Physics; Springer Berlin Heidelberg: Berlin, Heidelberg, 2010.
[51] G. D. Mahan, J. O. Sofo, *Proc. Natl. Acad. Sci.* **1996**, *93*, 7436.
[52] H. J. Xiang, D. J. Singh, *Phys. Rev. B* **2007**, *76*, 195111.
[53] X. Y. Huang, Z. Xu, L. D. Chen, *Solid State Commun.* **2004**, *130*, 181.
[54] X. Zhou, G. Wang, L. Zhang, H. Chi, X. Su, J. Sakamoto, C. Uher, *J Mater Chem* **2012**, *22*, 2958.
[55] M. S. Dresselhaus, G. Chen, M. Y. Tang, R. G. Yang, H. Lee, D. Z. Wang, Z. F. Ren, J.-P. Fleurial, P. Gogna, *Adv. Mater.* **2007**, *19*, 1043.
[56] Y.-L. Jin, X.-Z. Li, D. J. Sellmyer, *Mater. Charact.* **2018**, *136*, 302.